\documentclass[a4paper,10pt]{article} 
\usepackage[left=.65in,top=.65in,bottom=.65in,right=.65in]{geometry}
\usepackage{setspace} 
\usepackage{amsmath}
\usepackage{amsthm}
\usepackage{amssymb}
\usepackage{graphicx}
\usepackage{booktabs}
\usepackage{float}
\usepackage{longtable}
\usepackage{caption}
\usepackage[T1]{fontenc}
\usepackage{helvet}
  
\usepackage{orcidlink}  
\usepackage{xcolor}  
\usepackage{url}
\usepackage{hyperref} 
\usepackage{tcolorbox}
\makeatletter
\renewcommand{\maketitle}{\bgroup\setlength{\parindent}{0pt}
    \begin{Large}\@title\end{Large}\\
    \vspace{0.5em}\\
    \begin{normalsize}\@author\end{normalsize}\\
    \vspace{0.3em}\\
    \@date
\egroup} 
\makeatother

\usepackage{subcaption}

\usepackage{ragged2e}

\usepackage[backend=biber,style=numeric,sorting=none]{biblatex}
\usepackage[symbol]{footmisc}

\title{Large Language Models Achieve Gold Medal Performance at the International Olympiad on Astronomy \& Astrophysics (IOAA)}
\author{%
Lucas Carrit Delgado Pinheiro\orcidlink{0009-0008-9405-9682}$^{*,1}$, 
Ziru Chen\orcidlink{0009-0006-9805-1018}$^{*,2}$, 
Bruno Caixeta Piazza\orcidlink{0009-0005-2449-7554}$^{3}$, 
Ness Shroff\orcidlink{0000-0002-4606-6879}$^{1}$,
Yingbin Liang\orcidlink{0000-0002-8635-2992}$^{1}$,\\
Yuan-Sen Ting\orcidlink{0000-0001-5082-9536}$^{4,5}$,
Huan Sun\orcidlink{0000-0001-6436-4813}$^{2}$
}

\date{\footnotesize  
$^{1}$ Department of Electrical and Computer Engineering, The Ohio State University, Columbus, OH 43210, USA.\\
$^{2}$ Department of Computer Science and Engineering, The Ohio State University, Columbus, OH 43210, USA.\\
$^{3}$ Escola Politécnica, Universidade de São Paulo, São Paulo, SP 05508-010, Brazil.\\
$^{4}$ Department of Astronomy, The Ohio State University, Columbus, OH 43210, USA.\\
$^{5}$ Center for Cosmology and AstroParticle Physics (CCAPP), The Ohio State University, Columbus, OH 43210, USA. \\
$*$ Equal contribution.
}

\begin{document}
\maketitle
\thispagestyle{empty}

\begingroup\footnotesize
\vspace{.3cm}
\endgroup


\vspace{3em}
\small
\justifying

\noindent \textbf{While task-specific demonstrations show early success in applying large language models (LLMs) to automate some astronomical research tasks, they only provide incomplete views of all necessary capabilities in solving astronomy problems, calling for more thorough understanding of LLMs' strengths and limitations. So far, existing benchmarks and evaluations focus on simple question-answering that primarily tests astronomical knowledge and fails to evaluate the complex reasoning required for real-world research in the discipline. Here, we address this gap by systematically benchmarking five state-of-the-art LLMs on the International Olympiad on Astronomy and Astrophysics (IOAA) exams, which are designed to examine deep conceptual understanding, multi-step derivations, and multimodal analysis. With average scores of 85.6\% and 84.2\%, Gemini 2.5 Pro and GPT-5 (the two top-performing models) not only achieve gold medal level performance but also rank in the top two among $\sim$200--300 participants in all four IOAA theory exams evaluated (2022--2025). In comparison, results on the data analysis exams show more divergence. GPT-5 still excels in the exams with an 88.5\% average score, ranking top 10 among the participants in the four most recent IOAAs, while other models' performances drop to 48--76\%. Furthermore, our in-depth error analysis underscores conceptual reasoning, geometric reasoning, and spatial visualization (52--79\% accuracy) as consistent weaknesses among all LLMs. Hence, although LLMs approach peak human performance in theory exams, critical gaps must be addressed before they can serve as autonomous research agents in astronomy.}

\vspace{3em}

\section{Introduction}

With advancements in satellites and robotic telescopes, petabytes of new astronomical data are generated every year from surveys \cite{Ivezic2019,Gaia2023,DESI2024,Abdurrouf2022,Euclid2025}, transforming astronomy and astrophysics into a data-intensive subject. This vast volume of data has pushed the field beyond the limits of manual analysis, prompting efforts to develop autonomous approaches to augment and accelerate astronomical research. In response to this paradigm shift, earlier work applied neural networks and machine learning techniques to repetitive tasks beginning in the late 1980s \cite{Adorf1988,Angel1990,Odewahn1992,Storrie-Lombardi1992}. These methods have been used for sky object classification \cite{Kohonen1982,Mahonen1995,Brett2004}, anomaly detection \cite{Villar2021,Liang2023}, and other pattern recognition tasks in large astronomical datasets. However, these methods are not generalizable and can falter when moving across instruments, depths, noise regimes, or rare object types, let alone more challenging problems that demand advanced research skills, including complex computation, astronomical approximation, and conceptual reasoning. Automating scientific discovery in astronomy and astrophysics thus remains an ambitious challenge.

Recently, large language models (LLMs) are bringing new light to fulfill this ambition. Using natural language as the vehicle for reasoning \cite{Guo2025}, LLMs have demonstrated impressive problem-solving capabilities in several disciplines, such as biomedicine \cite{Zhou2024, Wang2025}, chemistry \cite{coscientist, yu-etal-2025-tooling}, and mathematics \cite{Trinh2024, lin2025goedelprover}. In astronomy and astrophysics, similar research efforts have also shown promising results in building autonomous AI agents with LLMs for some specific tasks, e.g., detecting gravitational-wave \cite{wang2025automatedalgorithmicdiscoverygravitationalwave} and interpreting multi-band galaxy observations \cite{sun2025interpretingmultibandgalaxyobservations}. Nonetheless, these task-specific demonstrations provide only a partial view of all capabilities required for astronomical research. Thus, it is necessary to examine and understand LLMs' strengths and weaknesses in solving astronomy problems through systematic benchmarking.

In this paper, we fill this gap by repurposing the International Olympiad on Astronomy and Astrophysics (IOAA) exams as a novel benchmark to comprehensively evaluate LLMs' performance in astronomical problem-solving. We believe IOAA problems are ideal testbeds for LLMs for three reasons. First, unlike existing benchmarks that merely test LLMs' astronomy knowledge through multiple choice, short answer, and true/false questions, such as AstroBench~\cite{ting2024astromlab1winsastronomy} from AstroMLab~\cite{dehaan2025astromlab4benchmarktoppingperformance} and Astro-QA~\cite{li2025}, IOAA exams are more ecologically valid because they evaluate the complex reasoning, creative problem-solving, and extended derivations required in actual astronomical research. Moreover, according to the official syllabus \cite{ioaa-syllabus}, IOAA problems cover a wide range of astronomical topics, including cosmology, spherical trigonometry, stellar astrophysics, celestial mechanics, photometry, and instrumentation, thus ensuring the comprehensiveness of our evaluation. Finally, IOAA integrates theoretical physics, observational constraints, and real-world astronomical data with mathematical computations, offering a unique assessment of LLMs' scientific problem-solving capabilities that complements existing evaluations on other international Olympiads, such as IMO \cite{huang2025gemini25procapable}, IPhO~\cite{qiu2025physicssupernovaaiagent}, and IOI~\cite{openai2025competitiveprogramminglargereasoning}.

We evaluate five state-of-the-art LLMs on the IOAA theory and data analysis exams from 2022 to 2025. Specifically, we select GPT-5, OpenAI o3, Gemini 2.5 Pro, Claude-4.1-Opus, and Claude-4-Sonnet, which are among the strongest models on AstroBench~\cite{ting2024astromlab1winsastronomy}, and possess necessary multimodal capabilities for IOAA problems. The observational exams in IOAA, requiring physical instruments and direct sky observations, are excluded from our evaluation due to LLMs' digital nature. All model outputs are independently graded by two IOAA experts following official rubrics.

\section{The International Olympiad on Astronomy and Astrophysics}

\paragraph{Exam Structure and Evaluation Scope.}
The International Olympiad on Astronomy and Astrophysics (IOAA) consists of three examination components: theory, data analysis, and observation, which are typically worth 300, 150, and 150 points, respectively. The only exception is the 2023 IOAA, in which the theory exam was worth 250 points, the other two components were worth 125 points. The scores in this study are expressed as percentages, which allows for fair comparisons between different editions of IOAA. When comparing the models' absolute scores in our analysis, we also normalize the 2023 exam scores to 300 for theory and 150 for data analysis. Due to the digital nature of LLMs, we only evaluate the theory and data analysis components, which do not require physical equipments such as telescopes or star charts. Although excluding the observation component, our setting retains substantial coverage of problem-solving capabilities from physical reasoning to statistical analysis. 

To comprehensively evaluate LLMs on these abilities, we compose an IOAA dataset that spans four years of IOAA exams (2022-2025), totaling 49 theory problems and 8 data analysis problems. Maintaining this diversity is crucial because, despite adherence to a standardized syllabus \cite{ioaa-syllabus}, each year’s problems are created by the host country’s committee, introducing natural variation in style, difficulty, and topic emphasis. By incorporating problems from multiple years, we strengthen our evaluation of model performance across such variations. To ensure clarity and reproducibility, each problem is distributed as a LaTeX file, often with embedded figures and plots that mirror the formats used in professional research.

\paragraph{Data Contamination Considerations.}
A critical advantage of our dataset is the inclusion of IOAA 2025 problems. Since the latest knowledge cutoff among all five evaluated LLMs is March 2025 (Claude Opus 4.1 and Claude Sonnet 4), the 2025 examination administered in August 2025 is naturally contamination-free. 

As aforementioned, we still include the 2022-2024 exams despite potential contamination to: (1) increase problem diversity across astronomical topics, (2) evaluate consistency of performance across years, and (3) assess contamination effects by comparing historical versus 2025 performance. As shown in Section \ref{sec:results} later, model scores on the uncontaminated 2025 exam closely match their average performance across all years, suggesting minimal contamination impact.

\paragraph{Theory Problem Characteristics and Categorization.}
The theory problems (5--75 points each) are designed to probe deep conceptual understanding through multi-step derivations, physical reasoning, and mathematical analysis. They cover a wide spectrum of topics --- including celestial mechanics, stellar astrophysics, cosmology, galactic dynamics, instrumentation, and observational astronomy --- and often require the integration of ideas across domains. For example, typical challenges might include combining spherical trigonometry with photometric principles for eclipse calculations, or linking cosmological models with observational constraints to test both theoretical knowledge and applied reasoning.  

We further distinguish theory problems by the type of reasoning they demand: About 37\% fall into Category I (Geometric/Spatial), which involves celestial sphere geometry, spherical trigonometry, and coordinate transformations. The remaining 63\% are Category II (Physics/Mathematics), emphasizing astrophysical calculations that do not rely on geometric visualization. This categorization, detailed in Appendix~\ref{appendixB}, allows us to better analyze the LLMs' performance and their specific incapabilities.

\paragraph{Data Analysis Problem Characteristics.}
Complementing theory questions, data analysis problems (45--105 points each) emphasize the practical skills central to modern astronomical research. Rather than abstract derivations, they require participants to work directly with observational evidence by extracting information from observational datasets, performing statistical analyses, interpreting plots and figures, and drawing scientific conclusions. These problems also use real-world astronomical data, such as light curves, spectra, stellar catalogs, and survey outputs, bridging theoretical knowledge with empirical inquiry.

\paragraph{Difficulty Distribution.}
We classify problem difficulty using human performance data. Based on median student scores, we define four levels: Easy (median above 50\%), Medium (median between 30\% and 50\%), Hard (median between 10\% and 30\%), and Extra Hard (median below 10\%). 
The resulting distribution in our dataset spans all difficulty levels, with 21\% Easy, 25\% Medium, 35\% Hard, and 19\% Extra Hard problems, taking into account both theory and data analysis exams.
This balance ensures that our evaluation captures both straightforward questions and the most challenging ones. 

\begin{table}[t]
\centering
\small
\caption{LLM Performance on IOAA theory and data analysis exams for different difficulty categories. All scores are normalized percentages of total points available.}
\label{tab:theory-da-summary}
\begin{tabular}{@{}lccccccccc@{}}
\toprule
 & \multicolumn{5}{c}{\textbf{Theory Exams}} & \multicolumn{4}{c}{\textbf{Data Analysis Exams}} \\
\cmidrule(lr){2-6} \cmidrule(lr){7-10}
\textbf{Model} & \textbf{Easy} & \textbf{Medium} & \textbf{Hard} & \textbf{Extra Hard} & \textbf{Overall} 
& \textbf{Easy} & \textbf{Medium} & \textbf{Hard} & \textbf{Overall} \\
 & & & & & Mean $\pm$ SD & & & & Mean $\pm$ SD \\
\midrule
GPT-5          & 84.6 & 76.6 & 91.5 & 77.2 & 84.2 $\pm$ 6.1
                & 100 & 75.5 & 93.1 & \textbf{88.5} $\pm$ 12.6\\
Gemini 2.5 Pro & 91.9 & 90.3 & 91.2 & 72.8 & \textbf{85.6} $\pm$ 8.0
                & 92.2 & 82.1 & 61.7 & 75.7 $\pm$ 15.5\\
OpenAI o3      & 97.0 & 84.5 & 78.8 & 63.1 & 77.5 $\pm$ 12.2
                & 91.6 & 58.8 & 62.4 & 67.7 $\pm$ 18.0\\
Claude Opus 4.1& 91.0 & 59.9 & 60.5 & 58.9 & 64.7 $\pm$ 13.5
                & 80.4 & 50.0 & 45.2 & 54.8 $\pm$ 19.1\\
Claude Sonnet 4& 79.2 & 62.0 & 57.5 & 54.7 & 60.6 $\pm$ 9.5
                & 75.1 & 35.5 & 43.6 & 47.9 $\pm$ 20.9\\
\bottomrule
\end{tabular}
\end{table}

\section{Results}
\label{sec:results}

\subsection{Performance of LLMs}

\paragraph{Theory Exams.}
As shown in Table~\ref{tab:theory-da-summary}, GPT-5 and Gemini 2.5 Pro are the strongest performers in theory exams, outperforming other models by 7 to 25 percentage points. Specifically, as shown in Table \ref{tab:ioaa-theory-ranks}, GPT-5 achieves the highest scores in 2022 (93.0\%), 2023 (89.6\%), and 2025 (86.8\%), while Gemini 2.5 Pro leads in 2024 (83.0\%). Overall, Gemini 2.5 Pro secures the best result (85.6\%) for its significantly better capabilities in solving geometric problems that dominate the 2024 exam, in which GPT-5 fails to obtain a high score. We include more detailed analysis of the models' capabilities and failure modes in Section~\ref{sec:theory-errors}.

Despite its overall strong performance, we have noticed that GPT-5 performs better on the hard questions than on the easy and medium ones. Our analysis indicates three reasons of this seemingly unusual performance oscillation. Firstly, the number of problems per difficulty level is small, which naturally allows for some variance in model performance. There are only 10 easy questions and 11 medium questions, worth a total of around 185 and 151 points, respectively, out of a total of 1200 for all categories. As a result, a few mistakes are already enough to cause a considerable variance on the performance of a model in each category. The second reason is that GPT-5 makes a number of significant mistakes on the 2024 exam, many of which come from problems involving tasks related to geometry and spatial visualization (more details in Section \ref{comp-human}). Finally, GPT-5 can occasionally make mistakes on astrophysics. For example, in question 9 from the 2024 exam, which is classified as easy, GPT-5 misses a total of 18 points due to the combination of a conceptual mistake and a miscalculation. This mistake alone accounts for almost 10\% of all points available for easy difficulty level. For these reasons, we suggest that there is no obvious misbehavior of GPT-5 that accounts for its lower performance on easy and medium questions. It is possible that a larger dataset can mitigate the effect of few occasional mistakes and result in a more balanced distribution across difficulty categories.

Other models also demonstrate competitive performance: OpenAI o3 scores 77.5\% overall and maintains a clear margin of 13--17 percentage points over the Claude models, where Claude Opus 4.1 scores 64.7\% and Claude Sonnet 4 scores 60.6\%. In addition, their performances decrease as the difficulty level increases. Although these three models perform comparably to each other and even achieve very positive scores on AstroMLab~\cite{dehaan2025astromlab4benchmarktoppingperformance}, a simpler benchmark with multiple-choice questions, our evaluation reveals substantial performance gaps. These results urge the need for more holistic evaluations of LLMs in astronomy to test the models' problem-solving capabilities beyond simple knowledge recalling.

\paragraph{Data Analysis Exams.}

While LLMs demonstrate near-peak performance in theory exams, data analysis exams speak more of their nuanced capabilities and limitations (Table~\ref{tab:theory-da-summary}). GPT-5 demonstrates exceptional data analysis capabilities with an 88.5\% overall score, which exceeds its theory exam performance (84.2\%). This improvement contrasts sharply with all other models, which show 10--15 percentage point drops from theory exams to data analysis exams. Such performance variances stem from the heavy reliance on plot interpretation and data visualization in data analysis exams. GPT-5's superior multimodal capabilities, evidenced by fewer plot reading and plotting errors in Figure~\ref{fig:error-analysis-da}, explain its dominance. To further advance LLMs in astrophysics, our results call for ecologically valid, multimodal benchmarks on astronomical data analysis, in addition to holistic evaluations. 

\subsection{Comparison to Human Performance}\label{comp-human}
To better understand LLMs' performance, we compare their scores against those of human participants according to IOAA's awarding criteria. 
In particular, the medals are awarded based on score distributions relative to the median performance (sum of the theory, data analysis, and observational exam scores): bronze for scores between 100\% and 130\% of the median, silver for 130\% to 160\% of the median, and gold for scores above 160\% of the median. 
Since our evaluation scope excludes the observation exam, we calculate medal thresholds for the theory exams and data analysis exams respectively.

In theory exams (Table~\ref{tab:ioaa-theory-ranks}), the LLMs mostly perform above the gold medal cutoff. The only exception is Claude Sonnet 4, which receives a silver medal in the 2023 exam. The models are also consistently able to rank among the top positions compared to their human counterparts. In the 2022, 2024, and 2025 exams, all models achieve a ranking in the top 12 out of around 200--300 participants per year. Most remarkably, GPT-5 outperforms the best IOAA student in 2022, 2023, and 2025, and Gemini 2.5 Pro achieves the same in 2022 and 2023. OpenAI o3 can also outperform the best student in the 2023 exam. In the worst case where Claude Opus 4.1 and Claude Sonnet 4 do not perform comparably to the top students in the 2023 exam, they still obtain scores considerably higher than the median, ranking in the 45$^{\text{th}}$ and 62$^{\text{nd}}$ places, respectively.


\begin{table}[H]
\centering
\caption{Comparison of LLM performance to human performance on IOAA theory exams (2022–2025). Medals are determined by performance relative to the human median (bronze: 100--130\%, silver: 130--160\%, gold: >160\%). All scores are normalized to percentages. Each model is ranked separately with respect to the students.}
\label{tab:ioaa-theory-ranks}

\begin{minipage}{0.48\linewidth}
\centering
\small
\textbf{IOAA 2025}\\[2pt]
\begin{tabular}{@{}l@{\hspace{4pt}}rrrr@{}}
\toprule
 & \textbf{Score} & \textbf{vs. Median} & \textbf{Rank} & \textbf{Medal} \\
\midrule
GPT-5           & 86.8 & 443\% & 1  & Gold \\
Gemini 2.5 Pro  & 81.2 & 414\% & 2  & Gold \\
OpenAI o3       & 73.8 & 377\% & 3  & Gold \\
Claude Opus 4.1 & 69.8 & 356\% & 6  & Gold \\
Claude Sonnet 4 & 60.2 & 307\% & 10 & Gold \\
\midrule
\textit{Thresholds:}\\
Gold   & 31.3 & 160\% & -- & -- \\
Silver & 25.5 & 130\% & -- & -- \\
Bronze & 19.6 & 100\% & -- & -- \\
\bottomrule
\end{tabular}
\end{minipage}\hfill
\begin{minipage}{0.48\linewidth}
\centering
\small
\textbf{IOAA 2024}\\[2pt]
\begin{tabular}{@{}l@{\hspace{4pt}}rrrr@{}}
\toprule
 & \textbf{Score} & \textbf{vs. Median} & \textbf{Rank} & \textbf{Medal} \\
\midrule
Gemini 2.5 Pro  & 83.0 & 323\% & 2  & Gold \\
GPT-5           & 67.5 & 263\% & 2  & Gold \\
OpenAI o3       & 67.3 & 262\% & 2  & Gold \\
Claude Sonnet 4 & 57.8 & 225\% & 10 & Gold \\
Claude Opus 4.1 & 57.2 & 223\% & 12 & Gold \\
\midrule
\textit{Thresholds:}\\
Gold   & 41.1 & 160\% & -- & -- \\
Silver & 33.3 & 130\% & -- & -- \\
Bronze & 25.7 & 100\% & -- & -- \\
\bottomrule
\end{tabular}
\end{minipage}

\vspace{1em}

\begin{minipage}{0.48\linewidth}
\centering
\small
\textbf{IOAA 2023}\\[2pt]
\begin{tabular}{@{}l@{\hspace{4pt}}rrrr@{}}
\toprule
 & \textbf{Score} & \textbf{vs. Median} & \textbf{Rank} & \textbf{Medal} \\
\midrule
GPT-5           & 89.6 & 232\% & 1  & Gold \\
OpenAI o3       & 88.8 & 230\% & 1  & Gold \\
Gemini 2.5 Pro  & 86.4 & 224\% & 1  & Gold \\
Claude Opus 4.1 & 62.8 & 163\% & 45 & Gold \\
Claude Sonnet 4 & 57.4 & 149\% & 62 & Silver \\
\midrule
\textit{Thresholds:}\\
Gold   & 62.0 & 160\% & -- & -- \\
Silver & 50.1 & 130\% & -- & -- \\
Bronze & 38.6 & 100\% & -- & -- \\
\bottomrule
\end{tabular}
\end{minipage}\hfill
\begin{minipage}{0.48\linewidth}
\centering
\small
\textbf{IOAA 2022}\\[2pt]
\begin{tabular}{@{}l@{\hspace{4pt}}rrrr@{}}
\toprule
 & \textbf{Score} & \textbf{vs. Median} & \textbf{Rank} & \textbf{Medal} \\
\midrule
GPT-5           & 93.0 & 306\% & 1 & Gold \\
Gemini 2.5 Pro  & 91.8 & 302\% & 1 & Gold \\
OpenAI o3       & 80.2 & 263\% & 3 & Gold \\
Claude Opus 4.1 & 68.8 & 226\% & 5 & Gold \\
Claude Sonnet 4 & 67.0 & 220\% & 6 & Gold \\
\midrule
\textit{Thresholds:}\\
Gold   & 48.7 & 160\% & -- & -- \\
Silver & 39.6 & 130\% & -- & -- \\
Bronze & 30.4 & 100\% & -- & -- \\
\bottomrule
\end{tabular}
\end{minipage}
\end{table}
\begin{table}[H]
\centering
\small
\caption{Comparison of LLM performance to human performance on IOAA data analysis exams (2022–2025). Medals are determined by performance relative to the human median (bronze: 100--130\%, silver: 130--160\%, gold: >160\%). All scores are normalized to percentages. Each model is ranked separately with respect to the students.}
\label{tab:ioaa-da-ranks}

\begin{minipage}{0.48\linewidth}
\centering
\small
\textbf{IOAA 2025}\\[2pt]
\begin{tabular}{@{}l@{\hspace{4pt}}rrrr@{}}
\toprule
 & \textbf{Score} & \textbf{vs. Median} & \textbf{Rank} & \textbf{Medal} \\
\midrule
GPT-5           & 72.3 & 204\% & 6   & Gold \\
Gemini 2.5 Pro  & 70.0 & 197\% & 10  & Gold \\
OpenAI o3       & 64.7 & 182\% & 21  & Gold \\
Claude Opus 4.1 & 41.0 & 115\% & 117 & Bronze \\
Claude Sonnet 4 & 30.0 & 84.4\%& 177 & None \\
\midrule
\textit{Thresholds:}\\
Gold   & 56.8 & 160\% & -- & -- \\
Silver & 46.2 & 130\% & -- & -- \\
Bronze & 35.5 & 100\% & -- & -- \\
\bottomrule
\end{tabular}
\end{minipage}\hfill
\begin{minipage}{0.48\linewidth}
\centering
\small
\textbf{IOAA 2024}\\[2pt]
\begin{tabular}{@{}l@{\hspace{4pt}}rrrr@{}}
\toprule
 & \textbf{Score} & \textbf{vs. Median} & \textbf{Rank} & \textbf{Medal} \\
\midrule
GPT-5           & 91.0 & 173\% & 10  & Gold \\
Gemini 2.5 Pro  & 84.7 & 161\% & 24  & Gold \\
OpenAI o3  & 78.3 & 150\% & 39  & Silver \\
Claude Opus 4.1 & 62.0 & 118\% & 85  & Bronze \\
Claude Sonnet 4 & 50.7 & 97\%  & 125 & None \\
\midrule
\textit{Thresholds:}\\
Gold   & 84.0 & 160\% & -- & -- \\
Silver & 68.3 & 130\% & -- & -- \\
Bronze & 52.5 & 100\% & -- & -- \\
\bottomrule
\end{tabular}
\end{minipage}

\vspace{1em}

\begin{minipage}{0.48\linewidth}
\centering
\small
\textbf{IOAA 2023}\\[2pt]
\begin{tabular}{@{}l@{\hspace{4pt}}rrrr@{}}
\toprule
 & \textbf{Score} & \textbf{vs. Median} & \textbf{Rank} & \textbf{Medal} \\
\midrule
GPT-5           & 100  & 250\% & 1  & Gold \\
Gemini 2.5 Pro  & 83.2 & 208\% & 3  & Gold \\
Claude Opus 4.1 & 71.2 & 179\% & 7  & Gold \\
Claude Sonnet 4 & 64.8 & 162\% & 11 & Gold \\
OpenAI o3       & 63.2 & 158\% & 14 & Silver \\
\midrule
\textit{Thresholds:}\\
Gold   & 64.0 & 160\% & -- & -- \\
Silver & 52.0 & 130\% & -- & -- \\
Bronze & 40.0 & 100\% & -- & -- \\
\bottomrule
\end{tabular}
\end{minipage}\hfill
\begin{minipage}{0.48\linewidth}
\centering
\small
\textbf{IOAA 2022}\\[2pt]
\begin{tabular}{@{}l@{\hspace{4pt}}rrrr@{}}
\toprule
 & \textbf{Score} & \textbf{vs. Median} & \textbf{Rank} & \textbf{Medal} \\
\midrule
GPT-5           & 90.7 & 329\% & 1  & Gold \\
Gemini 2.5 Pro  & 65.0 & 236\% & 7  & Gold \\
OpenAI o3       & 64.7 & 234\% & 9  & Gold \\
Claude Sonnet 4 & 46.0 & 167\% & 47 & Gold \\
Claude Opus 4.1 & 45.0 & 163\% & 51 & Gold \\
\midrule
\textit{Thresholds:}\\
Gold   & 44.2 & 160\% & -- & -- \\
Silver & 35.9 & 130\% & -- & -- \\
Bronze & 27.6 & 100\% & -- & -- \\
\bottomrule
\end{tabular}
\end{minipage}

\end{table}

In data analysis exams (Table~\ref{tab:ioaa-da-ranks}), there is a greater variance in the model performances. GPT-5 and Gemini 2.5 Pro are still in the gold medal range in all exams, with GPT-5 even surpassing the best student in 2022 and 2023. OpenAI o3 reaches the gold range in 2022 and 2025 and the silver range in 2023 and 2024. Claude Opus 4.1 and Claude Sonnet 4 show gold-level performances in 2022 and 2023 but receives bronze or no medal 2024 and 2025. We provide detailed analysis of these LLMs' failure modes in Section~\ref{performance-type}. 

\subsection{Error Analysis}
\label{performance-type}
\subsubsection{Failure Modes in Theory Exams}
\label{sec:theory-errors}

To have a better understanding of LLMs' strengths and limitations in astronomical problem-solving, we analyze the performance of LLMs in IOAA theory exams by different problem types. We categorize the theory problems into two groups based on our grading team's expert assessment: 
\begin{itemize}
\item \textbf{Category I (Geometric/Spatial):} Problems requiring spatial visualization, including celestial sphere, spherical trigonometry, timekeeping systems, and vector geometry.
\item \textbf{Category II (Physics/Mathematics):} Problems focused on cosmological and astrophysical calculations and celestial mechanics without geometric visualization requirements.
\end{itemize}

Although not exhaustive, this categorization (Table~\ref{tab:category-summary}) highlights a systematic performance gap: Models achieve strong results on Category II physics problems (67–91\%) but perform substantially worse on Category I geometric problems (49–78\%), a differential of 15–26 percentage points. The disparity is most pronounced in the 2024 exam where Category I problems dominate (see Appendix~\ref{appendixB}) --- only Gemini 2.5 Pro sustains a relatively high performance (74.7\%), while other models' performances fall to 35–59\%. Still, Gemini's performance in Category I problems is 12.7 percentage points lower than that of Category II problems (91.3\%).  

\begin{table}[t]
\centering
\small
\caption{Model performance by problem category on IOAA theory rounds. Category I comprises geometric/spatial problems; Category II comprises physics/mathematics problems. All scores are percentages.}
\label{tab:category-summary}
\begin{tabular}{@{}lccccc@{}}
\toprule
& \textbf{2022} & \textbf{2023} & \textbf{2024} & \textbf{2025} & \textbf{Overall}\\
& & & & & Mean $\pm$ SD\\
\midrule
\multicolumn{6}{@{}l}{\textbf{Category I: Geometric/Spatial Reasoning}}\\
\midrule
Gemini 2.5 Pro & 100.0 & 67.5 & 74.7 & 72.0 & 78.6 $\pm$ 14.6\\
GPT-5 & 99.5 & 73.5 & 58.7 & 72.7 & 76.1 $\pm$ 17.0\\
OpenAI o3 & 80.5 & 71.1 & 48.4 & 56.7 & 64.2 $\pm$ 14.4\\
Claude Opus 4.1 & 73.5 & 33.7 & 35.8 & 68.0 & 52.8 $\pm$ 20.9\\
Claude Sonnet 4 & 79.0 & 41.0 & 38.4 & 51.3 & 52.4 $\pm$ 18.6\\
\midrule
\multicolumn{6}{@{}l}{\textbf{Category II: Physics/Mathematics}}\\
\midrule
Gemini 2.5 Pro & 87.8 & 95.8 & 97.3 & 84.2 & 91.3 $\pm$ 6.3\\
GPT-5 & 89.8 & 97.6 & 82.7 & 91.6 & 90.4 $\pm$ 6.1\\
OpenAI o3 & 80.0 & 97.6 & 100.0 & 79.6 & 89.3 $\pm$ 11.0\\
Claude Opus 4.1 & 66.5 & 77.2 & 94.1 & 70.4 & 77.1 $\pm$ 12.2\\
Claude Sonnet 4 & 61.0 & 65.6 & 91.4 & 63.1 & 70.3 $\pm$ 14.2\\
\bottomrule
\end{tabular}
\end{table}

Why do LLMs fail in geometric problems? Through qualitative analysis, we find that LLMs are subject to some fundamental issues beyond mere calculation errors. First, the models struggle with spherical trigonometry at a conceptual level. For instance, GPT-5 would write spherical trigonometry equations that violates basic geometric principles and attempt angle calculations inconsistent with great circle geometry. Moreover, all models show confusion with timekeeping systems and choose between tropical and sidereal years incorrectly. Some of their solutions even implicitly treat calendar and tropical years as identical. Finally, LLMs to date can only reason in natural language but not visualize or sketch spatial representation during thinking, which brings a natural disadvantage compared to human participants. These failure modes highlight multimodal reasoning, especially spatial and temporal, as an important future direction to make LLMs better at astronomical problem-solving.

We additionally note that this conclusion does not contradict with the models' higher performance on Category I in the 2022 exam. As we can also find in our qualitative analysis, the 2022 exam has a reduced number of Category I problems (only 4 out of 13), and one of them can be directly solved through mathematical and physical relations, if the models are already familiar with the concept of retrograde motion. Besides, the only long problem among the four questions in 2022 is subject to data contamination, since it is based on an existing 1981 study on tadpole and horseshoe orbits~\cite{DERMOTT19811}. Consequently, the higher scores on Category I in the 2022 exam do not necessarily indicate that the models, in particular those of GPT-5 and Gemini 2.5 Pro, have excelled geometric reasoning.

\begin{figure}[t]
    \centering
    \begin{minipage}{\linewidth}
    \includegraphics[width=1.0\linewidth]{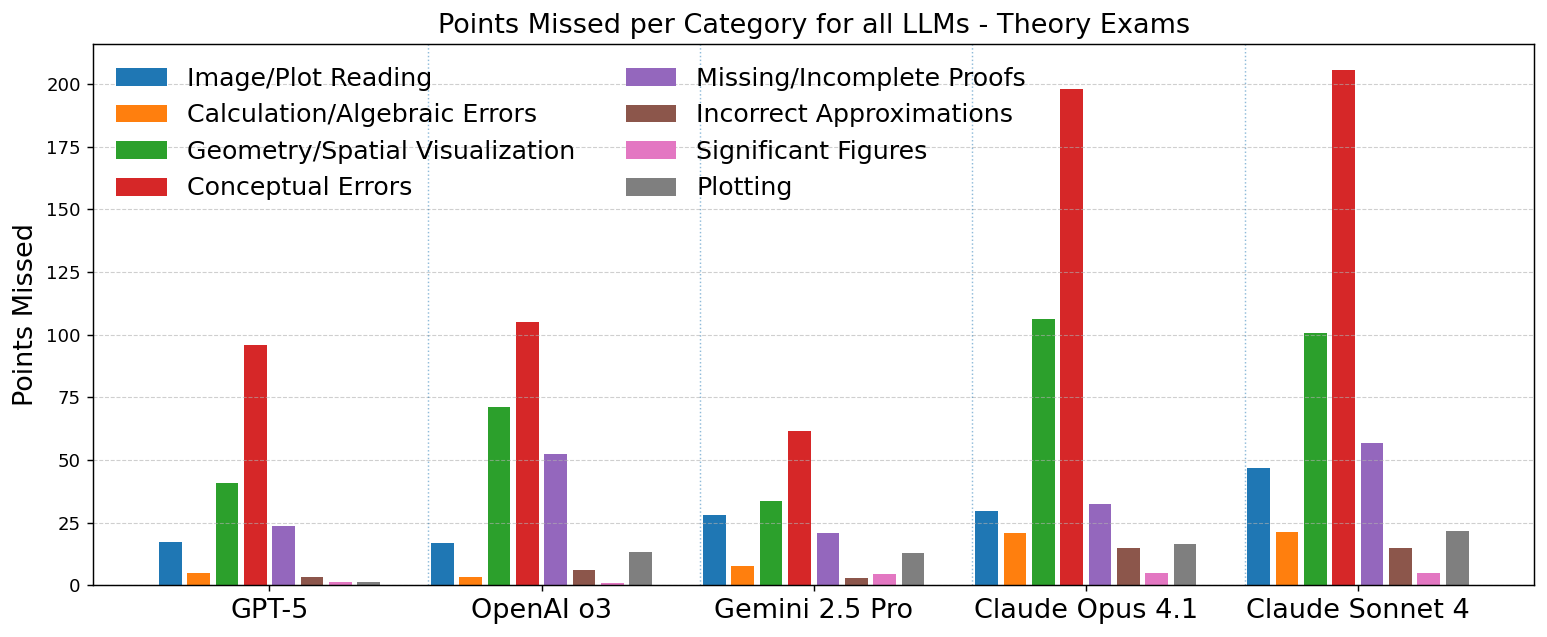}
    \subcaption{\label{fig:error-analysis-theory}}
    \end{minipage}
    \begin{minipage}{\linewidth}
    \includegraphics[width=1.0\linewidth]{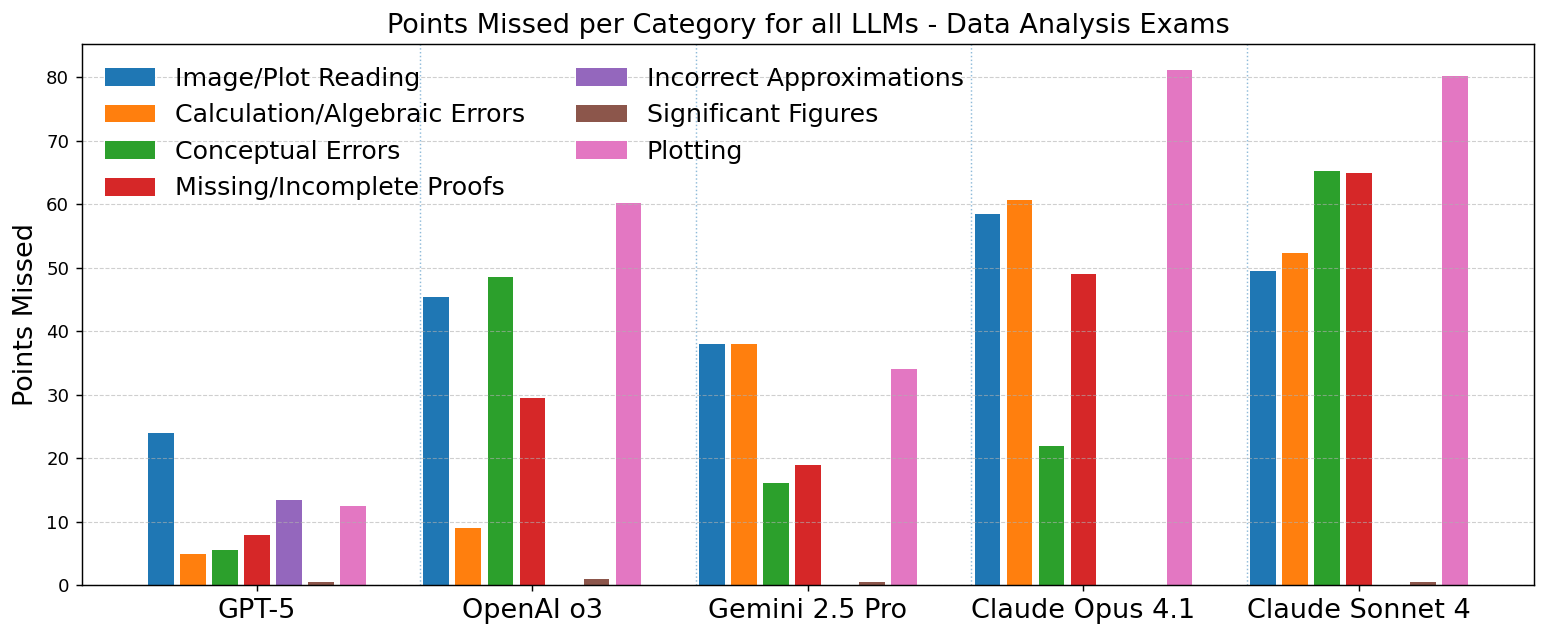}
    \subcaption{\label{fig:error-analysis-da}}
    \end{minipage}
    \caption{Distribution of points lost by error type across all models for (a) IOAA theory exams 2022--2025 (2023 scores normalized to 300 points) and (b) IOAA data analysis exams 2022--2025 (2023 scores normalized to 150 points). 
    In theory exams, conceptual errors and geometric/spatial visualization mistakes dominate across all models, together accounting for 60--70\% of total points lost. GPT-5 and Gemini 2.5 Pro show the lowest overall error rates, while Claude models show higher error rates. The distribution reveals that fundamental reasoning errors (conceptual and geometric) far outweigh computational mistakes, with the Claude models particularly struggling with conceptual understanding and all models except Gemini 2.5 Pro and GPT-5 showing notable geometric/spatial weaknesses.
    In data analysis exams, the error distribution is relatively balanced, with plotting as the most prominent category of errors for OpenAI o3, Claude Opus 4.1, and Claude Sonnet 4.}
\end{figure}

Besides qualitative analysis, we quantitatively classify all errors into eight categories to systematically identify LLMs' weaknesses. Figure~\ref{fig:error-analysis-theory} shows the distribution of points lost across these categories for all four theory exams. Conceptual errors --- incorrect approaches, misapplied formulas, and flawed reasoning --- are the most prevalent across all models, reflecting the fundamental challenge of achieving deep physical understanding. Unlike pure mathematics competitions such as the IMO, physics and astrophysics olympiads demand the integration of mathematical formalism with physical intuition, making them particularly valuable for assessing scientific reasoning abilities. Because such errors strike at the core of understanding, they typically occur across all problem types and result in severe point deductions.

The second-largest source of errors is geometric or spatial reasoning, concentrated entirely in Category I problems, which reinforces our finding that spatial reasoning represents a critical weakness of LLMs. Models frequently fail to visualize three-dimensional configurations, misidentify angles between celestial coordinates, or apply vector operations incorrectly in spherical geometry. These failures occur even when the geometry is described clearly in text, which is the case for the majority of the problems in Category I, suggesting that the limitation lies not only in multimodality but also in LLMs' fundamental capacity to approach tasks related to spatial reasoning.

Furthermore, astronomy olympiads place great emphasis on approximation and order-of-magnitude reasoning, given the vast scales involved. Although models generally handle approximations reasonably well, specific failures (see Appendix~\ref{appendixC}) highlight gaps in physical intuition. In particular, models often misjudge astronomical distances by orders of magnitude or fail to recognize when approximations are invalid under problem constraints.

Errors in interpreting plots and images, though confined to problems with visual inputs, also carry significant weight. This pattern aligns with known multimodal limitations in LLMs, such as chart understanding failures documented by \cite{wang2024charxivchartinggapsrealistic,yang2025effectivetrainingdatasynthesis}, and is consistent with Moravec’s Paradox: tasks simple for humans, such as visual interpretation, remain difficult for AI.  

Finally, missing or incomplete derivations are observed when models produce final expressions without showing intermediate steps, suggesting limits in mathematical reasoning transparency. The remaining categories, including calculation errors, notation precision, and approximation mistakes, result in minimal point loss, indicating decent computational skills.

\subsubsection{Failure Modes in Data Analysis Exams}
Unlike theory exams, the error distribution for data analysis exams (Figure \ref{fig:error-analysis-da}) is relatively even across multiple categories. As expected, plotting and plot and image reading are also causing point deductions in data analysis exams. The three less capable models, OpenAI o3, Claude Opus 4.1, and Claude Sonnet 4, have plotting as the main category of errors, while GPT-5 and Gemini 2.5 Pro have image and plot reading as the main source of point losses. 

Calculation errors are also responsible for a significant portion of points lost in data analysis exams, which is not the case for theory exams. For Gemini 2.5 Pro, calculation errors are even tied with image and plot reading as another major source of errors. This is because many data analysis problems involve long tables and require calculating of multiple values to generate a plot. Therefore, a large number of points is assigned to calculation steps on the official rubrics, which leads to higher chances of point losses when LLMs make calculation mistakes. It is also worth noting that conceptual mistakes and geometric errors, the main reasons of lost points in the theory exams, are not prominent in data analysis exams. While conceptual mistakes could happen on any problem and still contributes to point losses for most models in data analysis exams, the strong focus on plot reading and plotting tasks makes other types of errors more likely to occur. 

\section{Discussion and Conclusions}

In this paper, we present a comprehensive evaluation of state-of-the-art LLMs on the International Olympiad on Astronomy and Astrophysics (IOAA) examinations. Our evaluation demonstrates LLMs' remarkable capabilities in astronomical problem-solving. In the theory exams, all five models achieve gold medal performance, with GPT-5 and Gemini 2.5 Pro consistently ranking the first or the second among all human participants. In the data analysis exams, GPT-5 and Gemini 2.5 Pro also maintain their strong performance and reaches gold medal level, albeit with some reasonable rank drops. These achievements indicate that LLMs have obtained some genuine reasoning capabilities that rival talented human competitors in astronomy and astrophysics.

On one hand, our results suggest that LLMs are capable enough to serve as valuable AI co-scientists \cite{coscientist, gottweis2025aicoscientist} in some well-defined astronomical research tasks. For example, given their extraordinary performance in the theory exams, researchers may leverage LLMs to verify formulas, explore parameters, and cross-check astronomical concepts. On the other hand, LLMs are far from perfect to function as fully autonomous AI research agents. As shown in our analysis, their answers require careful validations to eliminate possible calculation errors and conceptual failures, such as violations of spherical trigonometry principles and incorrect temporal reasoning in astronomical contexts. Also, the considerable weaknesses in multimodal reasoning make them less favorable for data analysis.

Through benchmarking LLMs on the IOAA examinations, our study sheds light on their future development in astronomy and astrophysics. For instance, to improve LLMs' geometric and spatial reasoning, future work can implement visual sketchpad \cite{hu2024visual} so that the models can imitate humans to visualize or draw spatial representations and then find appropriate mathematical approaches to solve astronomical problems. Additionally, given the vast amount of astronomical data \cite{doi:10.1126/science.1170411}, we can synthesize visual question answering examples at scale to improve LLMs' multimodal understanding \cite{yang2025effectivetrainingdatasynthesis}. With focused development on these limitations and leveraging astronomy's collaborative research culture, we believe LLMs are poised to transition from impressive problem-solvers to valuable research partners, ultimately accelerating discovery in one of humanity's oldest sciences.

\section*{Methods}

\paragraph{LLM Evaluation.}

For each of the 57 IOAA problems (49 theory, 8 data analysis), we preprocess the LaTeX files by extracting figure references via regular expressions and encoding corresponding images as base64 binary data. Each model receives identical inputs consisting of: (1) the problem statement in LaTeX format, (2) embedded figures where applicable, (3) a standardized prompt template (Appendix~\ref{appendixA}, Table~\ref{tab:prompt}), and (4) a reference sheet containing 16 universal constants, 26 astronomical data values, and 6 calculus formulas—identical to materials provided to human participants.

The prompt template specified explicit requirements: complete step-by-step solutions with rigorous justification, proper LaTeX formatting with inline and display math modes, plot generation using tikzpicture and pgfplots packages, and document wrapping. Models were instructed to approach problems as astronomy experts participating in IOAA-level examinations.

We extract LaTeX code from model responses and compile them into PDF files. Both raw LaTeX and compiled PDFs are provided to graders, as compilation errors occasionally occur without affecting solution correctness. Each model is tested once per problem without retry attempts, reflecting examination conditions.

\paragraph{Human Grading Procedure.} 

Two expert graders have independently evaluated all model outputs: Lucas Carrit Delgado Pinheiro (IOAA participant 2018, team leader 2022, 2023, and 2025, and academic committee member 2024) and Bruno Caixeta Piazza (IOAA participant 2018, team leader 2021, academic committee member 2024, and observer 2025).

The grading procedure follows official IOAA rubrics with problems valued between 5-75 points (typically 300 points total for theory exams and 150 for data analysis). We follow the standard IOAA practice regarding point deductions and double penalty. In the case of a mistake that causes a numerical error to be carried forward to subsequent steps, we only deduct points from the step in which the mistake occurred. Answers with an unreasonable number of significant figures are only awarded half of the points assigned to the final answer.

For alternative solutions not matching official approaches, the graders apply equivalent standards: full credit for correct physics and mathematics regardless of method, partial credit proportional to progress toward solution, and consistent deductions for comparable errors across different approaches. Specific accommodations are made to ensure fair comparison with human participants who work with pencil and paper: accepting textual plot descriptions when LaTeX plotting code fails to compile (since humans draw by hand) and not penalizing LaTeX syntax errors that does not affect mathematical content (analogous to handwriting legibility).

These grading decisions mirror official IOAA procedures where human solutions receive credit for valid alternative approaches and partial credit for incomplete solutions. The accommodations compensate for the medium difference (LaTeX versus handwritten solutions) rather than lowering standards --- models are still required correct physics, complete derivations, and proper reasoning to earn points. This approach ensures that performance comparisons reflect genuine problem-solving capabilities rather than artifacts of the evaluation format.

After independent grading, the two graders have compared their scores for all 285 problem evaluations (57 problems × 5 models). All discrepancies are resolved through discussion until consensus is reached, with final scores representing agreed-upon evaluations. This dual-grader system from two experienced graders with consensus resolution ensures consistency and minimizes subjective bias in scoring.

\section*{Data Availability}
The IOAA 2025 problems and solutions can be found on the official IOAA 2025 website~\cite{ioaa2025-final-problems}. Problems and solutions from other years can be found on the resources section of the IOAA website~\cite{ioaa-past-problems}. The IOAA 2024 official website~\cite{ioaa2024-scores} provides a full list of participants' scores. Statistics for IOAA 2023 can be found on the proceedings for that year~\cite{ioaa-proceedings}. Our grading team has access to full score breakdowns for all IOAAs between 2022 and 2025 and consulted with a member of the IOAA Executive Committee before using these scores to generate the model rankings and medal estimates for this study.

\section*{Code Availability}
The code used in this work is publicly available from our GitHub repository at \url{https://github.com/OSU-NLP-Group/LLM-IOAA}.

\section*{Acknowledgements}
This research was sponsored in part by NSF CAREER \#1942980 and a gift from Amazon. Y.-S. Ting is supported by NSF \#AST-2406729. The views and conclusions contained herein are those of the authors and should not be interpreted as representing the official policies, either expressed or implied, of the U.S. government. The U.S. Government is authorized to reproduce and distribute reprints for Government purposes notwithstanding any copyright notice herein.

\section*{Author Contributions}
L.C.D. Pinheiro led the project, graded the outputs of the models, analyzed the scores of the models, wrote a significant portion of the manuscript, and generated most figures and tables. Z. Chen wrote the code to generate the models' responses to the IOAA exams, wrote portions of the manuscript, and revised the manuscript. B.C. Piazza graded the outputs of the models, analyzed the scores of the models, and revised the manuscript. N. Shroff and Y. Liang contributed to the formulation of the project and provided constructive feedback during weekly meetings. Y.-S. Ting is a subject matter expert who provided knowledge and guidance regarding the usage of LLMs in the field of astronomy, formulated the structure of the manuscript, provided constructive feedback during weekly meetings, and revised the manuscript. H. Sun conceived the original idea for the project, provided constructive feedback during weekly meetings, and revised the manuscript.

\section*{Competing Interests}
The authors declare that they have no competing interests.

\section*{Correspondence}
Correspondence should be addressed to \{carritdelgadopinheiro.1, chen.8336, ting.74, sun.397\}@osu.edu.

\printbibliography

\clearpage

\appendix

\section{LLM Prompting Strategy and Reference Materials}
\label{appendixA}

This appendix details the prompting methodology used to evaluate LLMs on IOAA problems. Table~\ref{tab:prompt} presents the complete prompt template designed to elicit rigorous, examination-style solutions from the models. The prompting strategy employs a two-part structure: (1) System/Developer instructions establishing the model's role as an astronomy expert, and (2) detailed User Message requirements specifying solution format, mathematical rigor, and output constraints.

The prompt explicitly requires step-by-step reasoning with complete justifications, proper LaTeX formatting for mathematical expressions, direct plot generation using tikzpicture and pgfplots packages, and document wrapping for automated extraction. These requirements mirror the expectations for human IOAA participants, ensuring comparable evaluation conditions. The warning that ``flawed or incomplete reasoning will receive no credit'' encourages comprehensive solutions rather than answer-focused responses.

Following the prompt template, we provide the standardized reference document supplied to all IOAA participants, containing essential constants and formulas. This document includes 16 universal constants (from Avogadro's number to particle masses), 26 astronomical parameters (solar and planetary properties, unit conversions, time definitions), and 6 fundamental calculus formulas. Providing these references ensures models are not penalized for constant recall while focusing evaluation on problem-solving capabilities.

\begin{table}[h]
\small
\centering
\caption{Prompt template for LLMs to solve IOAA problems and render their solutions into readable LaTeX formats.}
\begin{tabular}
{
@{\hspace{0pt}}ll@{\hspace{0pt}}
}
\toprule
\textbf{Message Roles} & \textbf{Instructions}\\
\midrule

System/Developer$^\dag$ & Ignore all previous instructions and DO NOT worry about fitting your answer in a single chat window.\\
& You are an expert in Astronomy and Astrophysics who is participating in an International Olympiad\\
& on Astronomy and Astrophysics (IOAA) level exam.\\

\midrule

User Message & Please think step by step and solve the given problem with a complete, detailed, and thorough\\
& answer.\\
& \\
& Please rigorously justify and clearly explain each step of your solution and do not skip\\
& important steps. You have unlimited space to write your answer. A correct final answer with \\
& flawed or incomplete reasoning will receive no credit.\\
& \\
& Please use LaTeX to clearly format your answer, especially properly wrapping math expressions \\
& in `\$...\$` for inline math and `\textbackslash[...\textbackslash]` for display math. A poorly formatted LaTeX solution\\
& that cannot compile will receive no credit.\\
& \\
& When asked to draw a plot, please use `tikzpicture` and `pgfplots` to directly make the figure\\ 
& in LaTeX and provide a clear and correct caption of your plot to explain your reasoning.\\
& A missing figure or poor caption will receive no credit.\\
& \\
& Please remember to wrap your solution in `\textbackslash begin\{document\}` and `\textbackslash end\{document\}`,\\
& as the grader will use them to extract the solution. A solution that cannot be extracted will\\
& receive no credit.\\
& \\
& Here is the problem statement:\\
& \{\{problem\_latex\}\}\\
& \\
& You may refer to the following document when solving the problem:\\
& \{\{constants\_and\_formulas\}\}$^\ddag$\\

\bottomrule
\end{tabular}
\caption*{ $^\dag$OpenAI now only supports ``Developer'' role as a replacement of the ``System'' role. $^\ddag$The document with useful constants and formulas is included in the next two pages. \label{tab:prompt}}
\end{table}

\begin{tcolorbox}[title=Document with useful constants and formulas used in our prompt.]
\label{toc}
\small

\textbf{Universal Constants}

\begin{tabular}{l l}
Avogadro constant & $6.022 \times 10^{23}\ \text{mol}^{-1}$ \\
Boltzmann constant & $1.381 \times 10^{-23}\ \text{J K}^{-1}$ \\
Charge of electron $e$ & $1.602 \times 10^{-19}\ \text{C}$ \\
Planck constant & $6.626 \times 10^{-34}\ \text{J s}$ \\
Speed of light in vacuum & $2.998 \times 10^{8}\ \text{m s}^{-1}$ \\
Universal gravitational constant & $6.674 \times 10^{-11}\ \text{N m}^2\ \text{kg}^{-2}$ \\
Universal gas constant & $8.315\ \text{J mol}^{-1}\ \text{K}^{-1}$ \\
Stefan-Boltzmann constant & $5.670 \times 10^{-8}\ \text{W m}^{-2}\ \text{K}^{-4}$ \\
Wien’s displacement constant & $2.898 \times 10^{-3}\ \text{m K}$ \\
Permittivity of free space & $8.854 \times 10^{-12}\ \text{m}^{-3}\ \text{kg}^{-1}\ \text{s}^4\ \text{A}^2$ \\
Permeability of free space & $1.257 \times 10^{-6}\ \text{N A}^{-2}$ \\
Mass of electron & $9.109 \times 10^{-31}\ \text{kg} = 0.511\ \text{MeV}/c^2$ \\
Mass of proton & $1.673 \times 10^{-27}\ \text{kg} = 938.272\ \text{MeV}/c^2$ \\
Mass of neutron & $1.675 \times 10^{-27}\ \text{kg} = 939.565\ \text{MeV}/c^2$ \\
Mass of deuteron & $3.344 \times 10^{-27}\ \text{kg} = 1875.613\ \text{MeV}/c^2$ \\
Mass of He nucleus & $6.645 \times 10^{-27}\ \text{kg} = 3727.181\ \text{MeV}/c^2$ \\
\end{tabular}

\vspace{1cm}

\textbf{Astronomical Data}

\begin{tabular}{l l}
Mass of Sun & $1.988 \times 10^{30}\ \text{kg}$ \\
Radius of Sun & $6.957 \times 10^{8}\ \text{m}$ \\
Luminosity of Sun & $3.828 \times 10^{26}\ \text{W}$ \\
Effective temperature of Sun & $5772\ \text{K}$ \\
Apparent magnitude of Sun (V-band) & $-26.74$ \\
Absolute magnitude of Sun (V-band) & $+4.82$ \\
Apparent bolometric magnitude of Sun & $-26.83$ \\
Absolute bolometric magnitude of Sun & $+4.74$ \\
Solar constant (above atmosphere of Earth) & $1361\ \text{W m}^{-2}$ \\
Apparent angular diameter of Sun (from Earth) & $\approx 32'$ \\
Mass of Earth & $5.972 \times 10^{24}\ \text{kg}$ \\
Radius of Earth & $6.378 \times 10^{6}\ \text{m}$ \\
Axial tilt of Earth & $23^\circ 26'$ \\
Inclination of lunar orbit to ecliptic & $5^\circ 8' 43''$ \\
Mass of Jupiter & $1.898 \times 10^{27}\ \text{kg}$ \\
Radius of Jupiter & $6.991 \times 10^{7}\ \text{m}$ \\
1 Astronomical Unit (au) & $1.496 \times 10^{11}\ \text{m}$ \\
1 parsec (pc) & $3.086 \times 10^{16}\ \text{m}$ \\
1 light-year (ly) & $9.461 \times 10^{15}\ \text{m}$ \\
1 jansky (Jy) & $10^{-26}\ \text{W m}^{-2}\ \text{Hz}^{-1}$ \\
1 tropical year & $365.2422\ \text{days} = 3.156 \times 10^7\ \text{s}$ \\
 & $= 365\ \text{d}\ 5\ \text{h}\ 48\ \text{min}\ 46\ \text{s}$ \\
1 sidereal year & $365.2564\ \text{days} = 3.156 \times 10^7\ \text{s}$ \\
 & $= 365\ \text{d}\ 6\ \text{h}\ 9\ \text{min}\ 13\ \text{s}$ \\
Rate of precession of Vernal Equinox & $1^\circ$ per 71.6 years \\
\end{tabular}
\end{tcolorbox}

\begin{tcolorbox}[title=Document with useful constants and formulas used in our prompt (continued).]
\small
\textbf{Calculus Related Formulas}

\[
\frac{d}{dx} x^n = n x^{n-1}
\]

\[
\frac{d}{dx} \sin(kx) = k \cos(kx)
\]

\[
\frac{d}{dx} \cos(kx) = -k \sin(kx)
\]

\[
\frac{d}{dx} \tan(kx) = k \sec^2(kx)
\]

\[
\int x^n dx = \frac{x^{n+1}}{n+1} + C, \quad n \neq -1
\]

\[
f(x) \approx f(x_0) + \left.\frac{df}{dx}\right|_{x=x_0} (x - x_0), \quad x \approx x_0
\]

\end{tcolorbox}

\section{Problem Categorization and Difficulty Analysis}
\label{appendixB}

This appendix provides a comprehensive breakdown of all IOAA problems used in our evaluation, categorized by difficulty level, problem type, and point value. The classification system serves two purposes: (1) understanding the distribution of problem types across examinations, and (2) contextualizing model performance relative to problem difficulty.

\subsection{Difficulty Classification Methodology}

We determine difficulty levels objectively based on student performance. Table~\ref{tab:diff-ranges} defines our classification system using the ratio of median student score to total available points. Problems where the median student achieves over 50\% of available points are classified as Easy, reflecting mastery by most participants. Medium difficulty problems (30-50\% median score) challenge students while remaining accessible to strong performers. Hard problems (10-30\% median) and Extra Hard problems (<10\% median) distinguish exceptional students, with the latter category representing problems that stump even most medal winners.

\begin{table}[H]
\centering
\caption{Difficulty classification based on ratio of median student score to total available points.}
\label{tab:diff-ranges}
\begin{tabular}{@{}cc@{}}
\toprule
\textbf{Median/Total Points} & \textbf{Difficulty Level}\\
\midrule
(0.50, 1.00] & Easy \\ 
(0.30, 0.50] & Medium \\
(0.10, 0.30] & Hard \\ 
$[$0.00, 0.10] & Extra Hard \\
\bottomrule
\end{tabular}
\end{table}

\subsection{Theory Problem Analysis}

Table~\ref{tab:theory-breakdown} presents all 49 theory problems across four years of IOAA examinations. Problems are further classified into Category I (Geometric/Spatial) and Category II (Physics/Mathematics) as discussed in the main text. Several patterns emerge from this classification:

\begin{longtable}{@{}clcrcc@{}}
\caption{Breakdown of IOAA theory questions by year, difficulty, and category.}
\label{tab:theory-breakdown}\\
\toprule
\textbf{Year} & \textbf{Q\#} & \textbf{Question Title} & \textbf{Points} & \textbf{Difficulty} & \textbf{Category}\\
\midrule
\endfirsthead
\multicolumn{6}{c}{\textit{Table \ref{tab:theory-breakdown} continued from previous page}}\\
\toprule
\textbf{Year} & \textbf{Q\#} & \textbf{Question Title} & \textbf{Points} & \textbf{Difficulty} & \textbf{Category}\\
\midrule
\endhead
\midrule
\multicolumn{6}{r}{\textit{Continued on next page}}\\
\endfoot
\bottomrule
\endlastfoot

\textbf{2022} & 1 & Planck's Units & 10 & Easy & II \\
 & 2 & Circumbinary Planet & 10 & Easy & II \\
 & 3 & Expanding Ring Nebula & 10 & Medium & II \\
 & 4 & Journey Between Galaxies & 10 & Hard & II \\
 & 5 & Flaring Protoplanetary Disk & 10 & Extra Hard & I \\
 & 6 & Photometry of Binary Stars & 20 & Medium & II \\
 & 7 & Georgia to Georgia & 20 & Hard & I \\
 & 8 & Ring of a Planet & 20 & Hard & II \\
 & 9 & Solar Retrograde Motion on Mercury & 20 & Extra Hard & I \\
 & 10 & Accretion & 20 & Hard & II \\
 & 11 & Dyson Sphere & 50 & Medium & II \\
 & 12 & Co-Orbital Satellites & 50 & Hard & I \\
 & 13 & Relativistic Beaming & 50 & Extra Hard & II \\
\midrule
\textbf{2023} & 1 & Neptune & 5 & Easy & I \\
 & 2 & Magnetic Field & 5 & Extra Hard & II \\
 & 3 & Microlensing & 5 & Easy & II \\
 & 4 & Europa & 10 & Medium & II \\
 & 5 & Dark Energy & 12 & Easy & II \\
 & 6 & Bolometer & 13 & Medium & I \\
 & 7 & Libration & 20 & Extra Hard & I \\
 & 8 & Neutrinos & 20 & Extra Hard & II \\
 & 9 & Second Eclipse & 20 & Medium & I \\
 & 10 & Aldebaran & 25 & Hard & I \\
 & 11 & X-Ray Emission from Galaxy Clusters & 30 & Easy & II \\
 & 12 & DART & 40 & Easy & II \\
 & 13 & LISA & 45 & Hard & II \\
\midrule
\textbf{2024} & 1 & Sundial & 10 & Easy & I \\
 & 2 & Galaxy Cluster & 10 & Easy & II \\
 & 3 & Asteroid & 10 & Hard & I \\
 & 4 & White Dwarf & 10 & Medium & II \\
 & 5 & CMB & 10 & Medium & II \\
 & 6 & Cluster Photography & 20 & Hard & II \\
 & 7 & Castaway & 20 & Hard & I \\
 & 8 & Binary Hardening & 25 & Hard & II \\
 & 9 & Physics of Accretion & 35 & Easy & II \\
 & 10 & Greatest Eclipse & 75 & Hard & I \\
 & 11 & Ground Tracks & 75 & Extra Hard & I\\
\midrule
\textbf{2025} & 1 & Daksha Mission & 10 & Medium & I \\
 & 2 & Makar-Sankranti & 10 & Medium & I \\
 & 3 & Gravitational Waves & 15 & Medium & II \\
 & 4 & Balmer Decrement & 15 & Hard & II \\
 & 5 & Quasars & 20 & Hard & II \\
 & 6 & Galactic Rotation & 20 & Extra Hard & I \\
 & 7 & Neutron Star Binary & 20 & Hard & II \\
 & 8 & Shadow of a Black Hole & 20 & Hard & II \\
 & 9 & Atmospheric Seeing & 35 & Extra Hard & I \\
 & 10 & Big Bang Nucleosynthesis & 35 & Extra Hard & II \\
 & 11 & Stars Through Graphs & 50 & Extra Hard & II \\
 & 12 & Hawking Radiation from Black Holes & 50 & Hard & II \\
\end{longtable}

\begin{itemize}
\item \textbf{Difficulty Distribution:} Across all years, problems span the full difficulty spectrum, with 20\% classified as Easy, 22\% as Medium, 35\% as Hard, and 22\% as Extra Hard. This distribution ensures comprehensive evaluation across skill levels.
\item \textbf{Category Balance:} Category I (geometric/spatial) problems constitute 37\% of problems and 39\% of total points, while Category II (physics/mathematics) represents 63\% of problems and 61\% of points. The 2024 exam was exceptional with Category I problems comprising 63\% of points.
\item \textbf{Point Weighting:} Problems range from 5 to 75 points, with higher-value problems ($\geq$35 points) predominantly falling in Hard or Extra Hard categories and testing integrated knowledge across multiple concepts.
\end{itemize}

\subsection{Data Analysis Problem Distribution}

Table~\ref{tab:da-breakdown} details the 8 data analysis problems evaluated. These problems uniformly test multimodal capabilities through plot interpretation, data extraction, and statistical analysis. Unlike theory problems, data analysis questions show less variation in structure but maintain difficulty diversity:

\begin{itemize}
\item \textbf{Difficulty Spread:} 25\% Easy, 37.5\% Medium, 37.5\% Hard, with no Extra Hard problems, reflecting the more constrained nature of data interpretation tasks.
\item \textbf{Point Values:} Data analysis problems carry substantial weight (45-105 points), emphasizing their importance in overall IOAA performance.
\item \textbf{Topic Coverage:} Problems span diverse astronomical applications from gravitational wave analysis to exoplanet detection, testing breadth of data interpretation skills.
\end{itemize}

\begin{longtable}{@{}clcrc@{}}
\caption{Breakdown of IOAA data analysis questions by year and difficulty.}
\label{tab:da-breakdown}\\
\toprule
\textbf{Year} & \textbf{Q\#} & \textbf{Question Title} & \textbf{Points} & \textbf{Difficulty}\\
\midrule
\endfirsthead
\multicolumn{5}{c}{\textit{Table \ref{tab:da-breakdown} continued from previous page}}\\
\toprule
\textbf{Year} & \textbf{Q\#} & \textbf{Question Title} & \textbf{Points} & \textbf{Difficulty}\\
\midrule
\endhead
\midrule
\multicolumn{5}{r}{\textit{Continued on next page}}\\
\endfoot
\bottomrule
\endlastfoot

\textbf{2022} & 1 & Gravitational Wave Astronomy & 45 & Medium\\
 & 2 & Galactic Surveys & 105 & Hard\\
\midrule
\textbf{2023} & 1 & Distance to the Large Magellanic Cloud & 50 & Easy\\
 & 2 & Isolated Black Hole & 75 & Hard\\
\midrule
\textbf{2024} & 1 & Photometric Comparison of Surveys & 75 & Easy\\
 & 2 & Shapley Hypothesis & 75 & Medium\\
\midrule
\textbf{2025} & 1 & 30 Years of Exoplanets & 90 & Medium\\
 & 2 & Predicting Arrival Times of CMEs & 60 & Hard\\
\end{longtable}

The classification system reveals that IOAA examinations provide balanced assessment across difficulty levels and problem types, making them suitable benchmarks for evaluating both strengths and limitations of LLM capabilities in astronomical problem-solving.

\section{Detailed Analysis of Model Failure Modes}
\label{appendixC}

This appendix provides concrete examples of systematic errors made by LLMs across different problem categories, illustrating the specific weaknesses identified in our quantitative analysis. These examples were selected to demonstrate recurring failure patterns rather than isolated mistakes, offering insights into fundamental architectural limitations that affect astronomical problem-solving capabilities.

The examples are organized by problem category and error type, with each case study including: (1) the problem context and correct solution approach, (2) specific mistakes made by different models, and (3) implications for the underlying capability gaps. These detailed analyses complement the aggregate statistics presented in the main text, providing qualitative understanding of how and why models fail on certain astronomical tasks.

\subsection{Category I Problems: Geometric and Spatial Reasoning Failures}

Category I problems reveal the most consistent and fundamental failures across all models. These examples demonstrate that spatial reasoning limitations extend beyond simple calculation errors to conceptual misunderstandings of three-dimensional geometry, coordinate systems, and astronomical reference frames.

\begin{itemize}

    \item \textbf{Example 1 - Eclipse Geometry (IOAA 2024 T10d):}

    This problem requires finding geographic coordinates for a solar eclipse center, testing understanding of three-dimensional celestial geometry. The correct solution involves vector operations recognizing that Sun, Moon, and Earth centers are generally non-collinear during eclipses.
    
    The solution requires finding scalar $k$ through:
    \[|\vec{\textbf{M}} + k\hat{\textbf{u}}|^{2} = R_{\oplus}^{2} \]

    Then determining the eclipse center position:
    \[\vec{\textbf{p}} = \vec{\textbf{M}} + k\hat{\textbf{u}}\]

    Only GPT-5 and Gemini 2.5 Pro correctly interpret the non-collinear geometry. OpenAI o3, Claude Opus 4.1, and Claude Sonnet 4 all incorrectly assume collinearity—a fundamental conceptual error that invalidates the entire solution approach. This failure demonstrates inability to visualize the actual three-dimensional configuration of celestial bodies during an eclipse, despite the geometry being fully described in text.

    \item \textbf{Example 2 - Spherical Trigonometry (IOAA 2024 T10h):}

    This problem requires extracting angle $\kappa$ from spherical triangles, as shown in Figure~\ref{fig:2024-Q10} to decompose vectors, testing mastery of spherical geometry principles.

    \begin{figure}[H]
        \centering
        \includegraphics[width=0.7\linewidth]{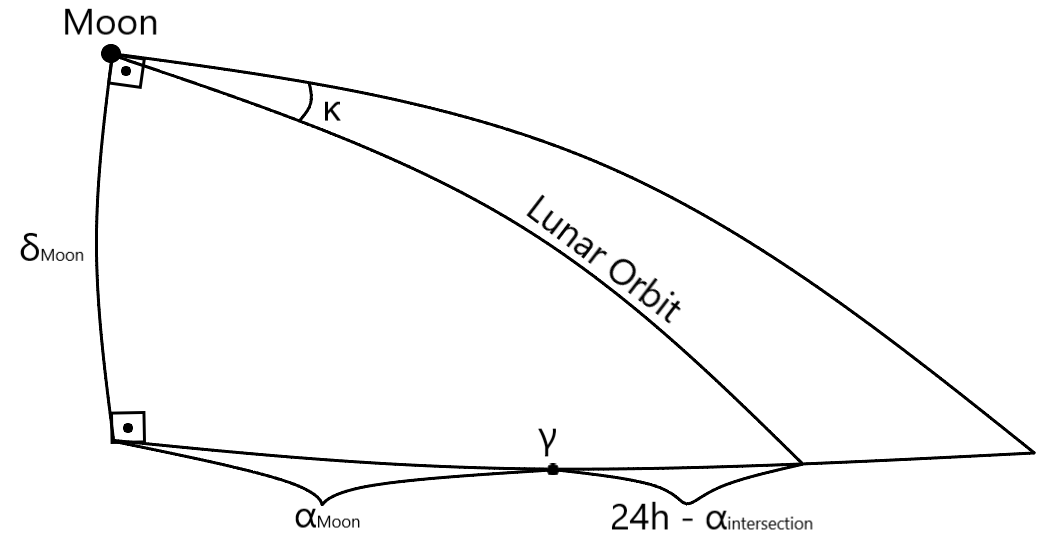}
        \caption{Spherical triangle configuration requiring angle extraction. Reproduced from the IOAA 2024 theory exam~\cite{ioaa-past-problems}.}
        \label{fig:2024-Q10}
    \end{figure}

    Gemini 2.5 Pro alone solves this correctly. GPT-5 and OpenAI o3 both pick incorrect angles for spherical trigonometry formulas. GPT-5 additionally claims two distinct angles were equal, contradicting basic spherical geometry principles. These errors reveal that models lack robust internal representations of non-Euclidean geometries essential for astronomical calculations.

    \item \textbf{Example 3 - Basic Angle Calculation (IOAA 2025 T01.1):}

    Even simple geometric tasks prove challenging. Given $\alpha = 120^\circ$ on Figure~\ref{fig:2025-Q1}, models need to find the angle between the $y$-axis and the normal to $D_2$.

    \begin{figure}[H]
        \centering
        \includegraphics[width=0.6\linewidth]{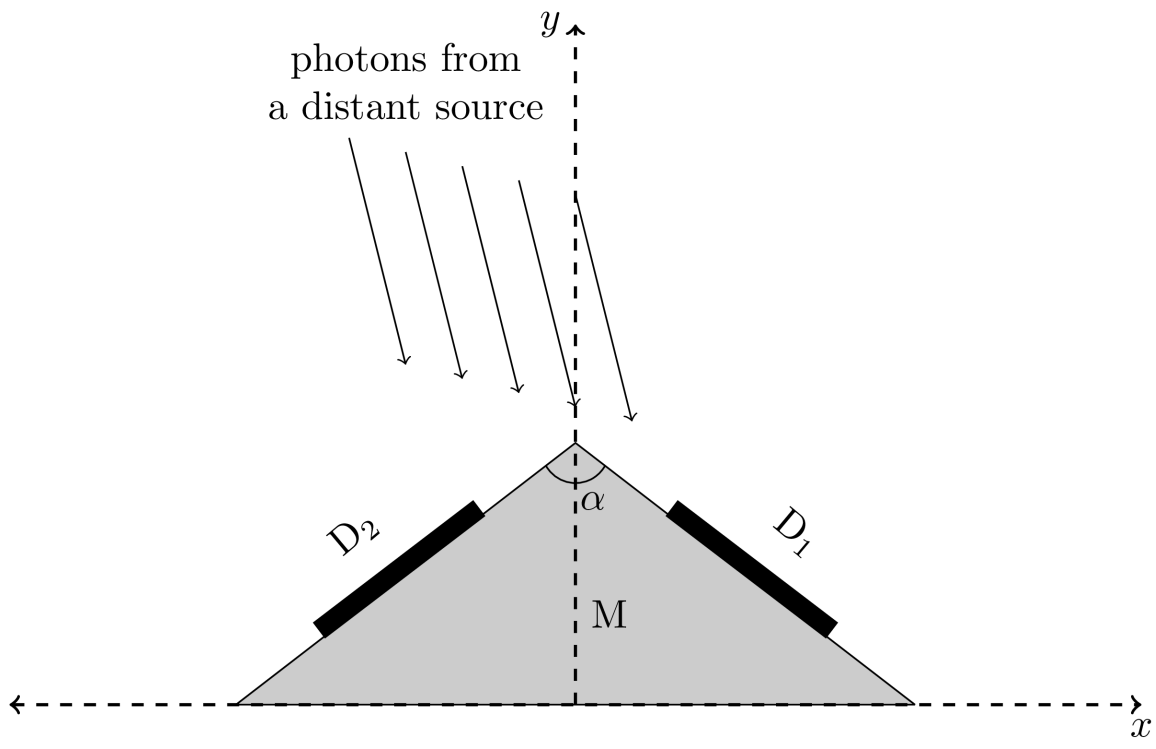}
        \caption{Detector configuration with $\alpha = 120^\circ$. Reproduced from the IOAA 2025 theory exam~\cite{ioaa2025-final-problems}.}
        \label{fig:2025-Q1}
    \end{figure}

    The correct answer is 30$^\circ$, obtainable through basic angle relationships. All models except for Gemini 2.5 Pro incorrectly calculate 60$^\circ$, despite the geometry being explicitly described in text. This failure on elementary geometry suggests systematic issues with angle visualization and spatial relationships.

    \item \textbf{Example 4 - Temporal Reference Frames (IOAA 2025 T02):}

    This problem tests the understanding of different astronomical year definitions and precession. For T02.1, no model correctly chooses between tropical and sidereal years for the required calculation. For T02.2, only Gemini 2.5 Pro succeeds; other models implicitly equate calendar and tropical years—a fundamental error that ignores leap year corrections. These failures indicate confusion about astronomical time systems and reference frame transformations.
    
\end{itemize}

\subsection{Category II Problems: Physics and Mathematical Reasoning}

While models perform better on Category II problems overall, specific examples reveal important limitations in physical reasoning and mathematical rigor.

\begin{itemize}
    \item \textbf{Example 1 - Temperature Estimation (IOAA 2022 T11 e):}

    The task requires students to estimate Earth's temperature in a hypothetical scenario in which a Dyson sphere completely blocks all incoming solar radiation. The models struggle with this problem and fail to account for the fact that the Dyson sphere itself would heat up due to absorbed solar energy. GPT-5 and Claude Opus 4.1 assume Earth's new temperature to be $0\,\mathrm{K}$, whereas Gemini 2.5 Pro and Claude Sonnet 4 defines an internal energy source within Earth to compute a nonzero equilibrium temperature.

    \item \textbf{Example 2 - Dyson Sphere Search (IOAA 2022 T11 i):}

    This question involves the estimation of the range of wavelengths in which we should search for a Dyson sphere built by a civilization in a distant galaxy. Despite this clear hint, all models -- except Gemini 2.5 Pro -- fail to account for the cosmological redshift when converting the temperature of the Dyson sphere into the expected observed wavelength.
\end{itemize}

\subsection{Multimodal Processing Failures}

\subsubsection{Plot and Image Reading Errors}

These examples demonstrate that multimodal limitations extend beyond simple OCR-like tasks to fundamental challenges in extracting quantitative information from scientific visualizations.

\begin{itemize}
    \item \textbf{Example 1 - Distance Measurements (IOAA 2025 T05.1):}

    This problem requires models to measure distances between reference marks using scale bar provided on Figure \ref{fig:2025-Q6}.

    \begin{figure}[H]
        \centering
        \includegraphics[width=0.32\linewidth]{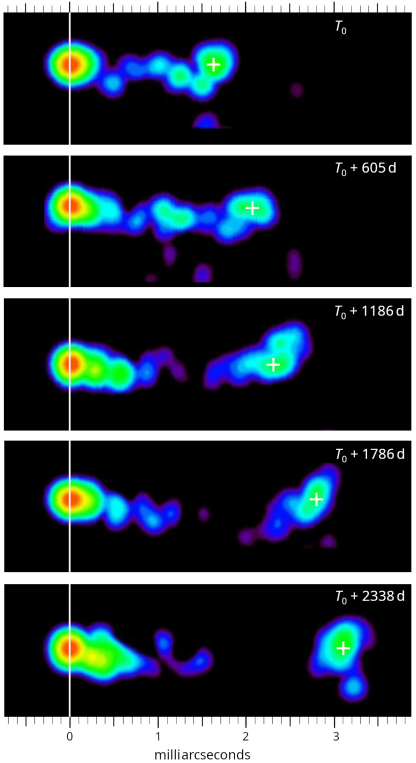}
        \caption{Quasar images requiring distance measurements between white line and plus markers. Reproduced from the IOAA 2025 theory exam~\cite{ioaa2025-final-problems}.}
        \label{fig:2025-Q6}
    \end{figure}

    Only GPT-5 obtains mostly correct measurements. Other models show errors ranging from 20-50\% despite understanding the task requirements. This reveals limitations in precise spatial measurement from images, critical for analyzing astronomical observations.

    \item \textbf{Example 2 - Blackbody Curve Identification (IOAA 2025 T10.2b):}

    Models demonstrate theoretical understanding of Wien's law and Stefan-Boltzmann relationship but fail to identify the correct plot out of the options displayed on Figure~\ref{fig:2025-Q10}.

    \begin{figure}[H]
        \centering
        \includegraphics[width=0.9\linewidth]{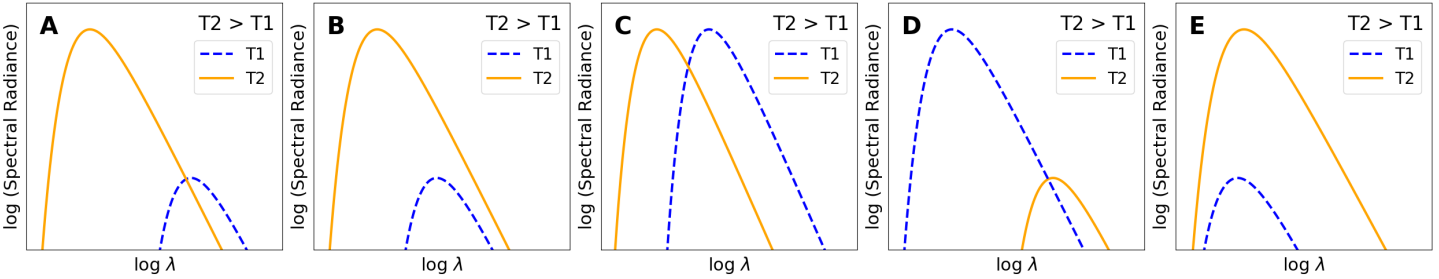}
        \caption{Multiple blackbody radiation curves requiring identification of correct temperature relationship. Reproduced from the IOAA 2025 theory exam~\cite{ioaa2025-final-problems}.}
        \label{fig:2025-Q10}
    \end{figure}

    All models articulate that higher temperature curves must peak at shorter wavelengths with higher overall intensity, yet none select the correct graph. This disconnect between conceptual understanding and visual pattern recognition highlights the challenge of integrating theoretical knowledge with visual analysis.

    \item \textbf{Example 3 - Complex Data Extraction (IOAA 2025 T12.2b):}

    This problem requires extracting specific values from the cosmological scale factor evolution plot displayed on Figure~\ref{fig:2025-Q12}.

    \begin{figure}[H]
        \centering
        \includegraphics[width=0.50\linewidth]{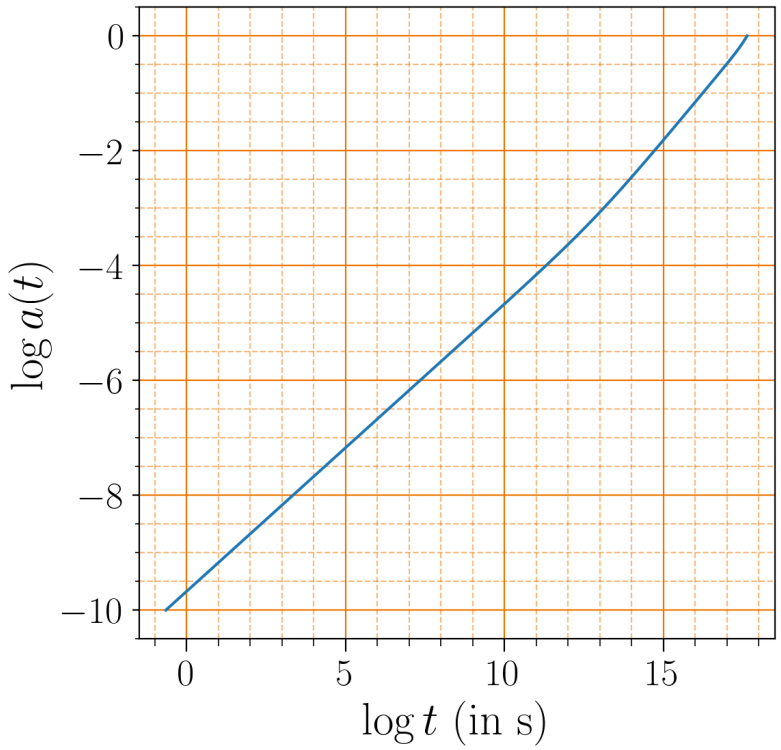}
        \caption{Scale factor evolution requiring value extraction for black hole mass calculations. Reproduced from the IOAA 2025 theory exam~\cite{ioaa2025-final-problems}.}
        \label{fig:2025-Q12}
    \end{figure}

    The complex reasoning chain—using extracted values to determine black hole mass ranges—causes most models to abandon the plot entirely, attempting alternative (incorrect) approaches. Only GPT-5 successfully integrates plot reading with physical reasoning, demonstrating that multimodal challenges compound when combined with complex problem-solving.
    
\end{itemize}

\subsection{Approximation and Mathematical Rigor}

\begin{itemize}
    \item \textbf{Example 1 - Inappropriate Approximations (IOAA 2025 T06.2):}

    All models incorrectly apply Oort's approximation for galactic rotation when an exact solution was derivable from given information. The problem statement's inclusion of this approximation for a subsequent part likely triggers inappropriate pattern matching, revealing over-reliance on common approximations without assessing their validity.

    \item \textbf{Example 2 - Selective Small-Angle Approximations (IOAA 2025 T09.2b):}

    In this Snell's law problem, models need to apply small-angle approximations selectively. Most models either apply the approximation universally (incorrect) or avoid it entirely (also incorrect). Only Gemini 2.5 Pro correctly identifies which angles permitted approximation, demonstrating the nuanced physical reasoning required for appropriate approximation use.

\end{itemize}

\subsection{Proof Completeness and Mathematical Communication}

\begin{itemize}

    \item \textbf{Example 1 - Incomplete Orbital Mechanics Derivation (IOAA 2024 T11g):}

    This problem requires deriving the time a satellite spends in the northern hemisphere. Claude Opus 4.1 establishes incorrect geometric relationships between eccentric and true anomalies but then skips to the "correct" final expression without justification. Claude Sonnet 4 simply states "After working through the integration (which involves elliptic integrals)" without providing any actual integration. These examples reveal tendencies to produce answer-focused responses rather than rigorous derivations, potentially reflecting training biases toward final answers over complete mathematical reasoning.

\end{itemize}

These detailed examples demonstrate that LLM failures in astronomical problem-solving stem from fundamental architectural limitations rather than simple knowledge gaps. The consistency of certain error types across models suggests these represent current technological boundaries rather than model-specific weaknesses. Understanding these failure modes is essential for both improving model architectures and identifying appropriate deployment contexts for current capabilities.

\newpage
\end{document}